\documentclass[aps,twocolumn,preprintnumbers,floatfix,nofootinbib,showpacs]{revtex4}

\usepackage{graphicx}
\usepackage{dcolumn}
\usepackage{bm}
\usepackage{epsfig}
\usepackage{color}

\usepackage{amsmath}
\usepackage{amssymb}

\def\fun#1#2{\lower3.6pt\vbox{\baselineskip0pt\lineskip.9pt
  \ialign{$\mathsurround=0pt#1\hfil##\hfil$\crcr#2\crcr\sim\crcr}}}

\input epsf

\newcommand{\be}{\begin{equation}}
\newcommand{\ee}{\end{equation}}
\newcommand{\bea}{\begin{eqnarray}}
\newcommand{\eea}{\end{eqnarray}}

\begin{document}



\title{Landau-Yang Theorem and Decays of a $Z'$ Boson into Two $Z$ Bosons}
\author{Wai-Yee Keung$^a$, Ian Low$^{b,c}$, Jing Shu$^{c,d}$}
\affiliation{
$^a$ Department of Physics, University of Illinois, Chicago, IL 60607 \\
$^b$ \mbox{Department of Physics and Astronomy, Northwestern University, Evanston, IL 60208} \\
$^c$ Theory Group, HEP Division, Argonne National Laboratory, Argonne, IL 60439\\
$^d$ \mbox{Enrico Fermi Institute and Department of Physics, University of
Chicago, Chicago, IL 60637} }

\begin{abstract}
We study the decay of a $Z'$ boson into two $Z$ bosons by extending
the Landau-Yang theorem to a parent particle decaying into two $Z$
bosons. For a spin-1 parent the theorem predicts: 1) there
are only two possible couplings and 2)
the normalized differential cross-section 
depends on kinematics only through a phase shift in the azimuthal angle between the two
decay planes of the $Z$ boson. 
When the parent is a $Z'$ 
the two possible couplings are anomaly-induced and CP-violating, 
respectively. 
{ At the CERN Large Hadron Collider their 
effects}  could be disentangled when both $Z$ bosons decay leptonically.

\end{abstract}


\maketitle


\noindent {\em Introduction -- }
A heavy $Z^\prime$ boson is ubiquitous in extensions of the 
Standard Model (SM) \cite{Langacker:2008yv}, 
which frequently modify the internal and/or spacetime symmetry
of the SM. The $Z'$ could be either the gauge boson of an extra $U(1)'$ group (embedded in a larger gauge group) or 
the Kaluza-Klein (KK) partner of the SM $Z$ boson propagating in extra dimensions.
 Its existence, if verified experimentally, has
important implications beyond the mere observation of
a new vector gauge boson. 
{ 
For example, if the $Z'$ belongs to an extra $U(1)'$ gauge group, 
there must be other 
degree of freedom responsible for giving it a mass. Furthermore, constraints from cancellation of anomalies
associated with the $U(1)'$ often require introducing new particles. On the other hand, if the $Z'$ is
the first KK mode of the SM $Z$ boson, then additional KK modes such as the partner of other SM gauge bosons
should be present as well. In other words, the discovery of a $Z'$ boson would point us to more 
beyond-the-Standard-Model (BSM) particles and interactions.
}

Here we wish to study the coupling of a new $Z'$ boson with two SM $Z$ bosons. Even though the mass
and couplings of a $Z'$ could span over a wide range of parameter space, we are mainly interested
in scenarios where the $Z'$ couples to SM only through gauge boson couplings. A well-motivated
example of such a possibility is in some versions of little Higgs theory with 
T parity \cite{Cheng:2003ju}. To proceed, we first
extend the Landau-Yang theorem \cite{Yang:1950rg} to decays of a parent particle into two $Z$ bosons.
The original theorem prohibits decays of a spin-1 particle into two photons. In the case of a spin-1
parent decaying into $ZZ$
final state, simple symmetry arguments allow us to re-derive the known fact 
that there are only two
possible couplings, which are similar to the anomalous couplings of triple neutral gauge bosons in the 
SM \cite{Gaemers:1978hg,Hagiwara:1986vm}. In addition,
we will show that the normalized differential cross section depends on kinematic
variables only through a phase shift
in the azimuthal angle between the two decay planes of the $Z$.



At the level of effective Lagrangian, the two operators contributing to the $Z'ZZ$ amplitude 
are both very interesting. One of
them, $O_{A}$, is induced by the anomaly, if exists, associated with the $U(1)'$. In so-called 
string-inspired models, $O_{A}$ is present when the $U(1)'$ anomaly is cancelled either
by heavy exotic fermions \cite{Chang:1988fq} or by an axion through 
the Green-Schwarz mechanism \cite{Anastasopoulos:2006cz}. In a
general non-linear sigma model (nl$\sigma$m), 
$O_A$ can arise from the Wess-Zumino-Witten (WZW) term as 
emphasized in \cite{Hill:2007nz}.  
{ (One cannot resist emphasizing the obvious:
the existence of WZW term depends on the UV completion of the nl$\sigma$m; see 
Ref.~\cite{Krohn:2008ye} for 
realizations with unbroken T parity.)} 
The second operator, $O_{CPV}$, on the other hand is CP-violating and
could arise from a {\em scalar} triangle loop in a two-Higgs-doublet
model \cite{Chang:1994cs}.  ($O_A$ is P-violating, but CP-conserving.)
It is possible that $O_A$ and $O_{CPV}$
interfere with each other, leading to potentially large CP-violating effects. 

Using results from the generalized Landau-Yang theorem, it is possible to
disentangle the effects of $O_A$ and $O_{CPV}$ 
in experiments by measuring the phase shift in the azimuthal angle of
$Z'\to ZZ \to 4\ell$. 
(Previous studies focus on only one operator, but not the presence of both. 
For example,
see Refs.~\cite{Barger:2007df,Kumar:2007zza} for 
studies on $O_A$ alone.) Since the
method does not rely on knowing the incoming beam axis, it allows us to consider general
cases when the $Z'$ is not directly produced but is part of a long decay-chain. 
Experimentally, the four leptons final state is very distinct and
well-studied since Higgs { boson} 
$\to ZZ \to 4\ell$ is one of the golden channels for Higgs
{ boson}
discovery.



\noindent {\em Generalized Landau-Yang Theorem --} 
The Landau-Yang theorem uses general principle of rotation invariance and space inversion to derive
certain selection rules governing decays of a parent particle into two photons. For a parent with 
spin less than two, the polarization state of the two photons 
{ is } completely fixed by the selection 
rule. When it comes to $ZZ$ final state, the only difference is the $Z$ boson has one extra 
polarization state than the photon due to its massive nature. We will choose the coordinate system to
be the rest frame of the parent particle in which $Z_1$ moves in the $+z$ direction and 
$Z_2$ in the $-z$ direction. (See Fig.~1.) Secondary decays of the $Z$ boson are described in 
the rest frames of $Z_1$ and $Z_2$, obtained by 
boosting along the $+z$ and $-z$ directions, respectively.
In our coordinate system, the helicity state of the two $Z$'s are defined
as
\begin{eqnarray}
\epsilon^{(1)}_0 &=& \gamma(\beta, 0,0,1) = \frac{m_Y}{2m_Z}(\beta,0,0,1), \\
\epsilon^{(2)}_0 &=& \gamma(-\beta, 0,0,1) = \frac{m_Y}{2m_Z}(-\beta,0,0,1), \\
\epsilon^{(1)}_{\pm} &=&  (0,\mp1,-i,0)/\sqrt{2} = \epsilon^{(2)}_{\mp},
\end{eqnarray}
where $m_Y$ is the mass of the parent particle $Y$.
We follow closely the notation and convention of \cite{Hagiwara:1986vm} where $\epsilon_{0123}
=-\epsilon^{0123}=1$. Also notice that we have chosen both of the longitudinal polarizations
to be along $+z$, even though the direction of motion is opposite. In the end there are nine possible
polarization states for the $ZZ$, $\Psi^{\lambda_1 \lambda_2}$, where $\lambda_{1,2} =+,-, 0$.

With the above definitions, it is straightforward to work out the action on 
$\Psi^{\lambda_1 \lambda_2}$ under the following three symmetry transformations: 1) ${\cal R}^{\psi}$ 
is the rotation 
around the $z$ axis by an angle $\psi$, 2) ${\cal R}^{\xi}$ is the rotation around 
the $x$ axis by $\pi$, and 3) P is the space inversion. In effect ${\cal R}^{\psi}$ imposes
angular momentum conservation along the $z$ axis and ${\cal R}^{\xi}$ enforces the Bose symmetry.
The selection rule for a parent with spin $J\le 1$ and 
parity P$=\pm$ is summarized in Table I.
\begin{table}[b]
\caption{\label{table1}Helicity states $\Psi^{\lambda_1\lambda_2}$ of the $Z$ bosons}
\begin{ruledtabular}
\begin{tabular}[b]{c|cc}
   P $\backslash J$     & $0$      & $1$ \\
\hline
$+$ & $\Psi^{++}+\Psi^{--}, \Psi^{00}$ &    $\Psi^{+0}-\Psi^{0-}, \Psi^{0+}-\Psi^{-0}$      \\
\hline 
$- $ & $\Psi^{++}-\Psi^{--}$   & $\Psi^{+0}+\Psi^{0-}, \Psi^{0+}+\Psi^{-0}$
\end{tabular}
\end{ruledtabular}
\end{table}

We would like to focus on a $J=1$ parent particle and denote the helicity amplitude 
of $Y(\kappa)\to Z_1(\lambda_1)Z_2(\lambda_2)$ as
${\cal M}_{\kappa,\lambda_1\lambda_2}$, where $\kappa$ is the spin projection of the parent
along the $+z$ axis. (We define $\epsilon^{(Y)}_\pm = \epsilon^{(1)}_\pm$ and 
$\epsilon_0^{(Y)} = (0,0,0,1)$.)
Immediately we see that the only non-vanishing amplitudes are
${\cal M}_{+,+0},\ {\cal M}_{+,0-},\ {\cal M}_{-,-0}$, and ${\cal M}_{-,0+}$. In particular,
amplitudes ${\cal M}_{0,\pm\pm}$ and ${\cal M}_{0,00}$ 
are forbidden by the Bose symmetry, whereas all others vanish due to angular
momentum conservation. It is interesting to further consider the action of the non-vanishing
amplitudes under $R^\xi$ and P:
\begin{eqnarray}
\label{rxi}
{\cal R}^\xi&:& {\cal M}_{+,+0} \leftrightarrow -{\cal M}_{-,0+}, \quad
   {\cal M}_{+,0-} \leftrightarrow - {\cal M}_{-,-0}; \\
\label{rp}
\mbox{P}&:& {\cal M}_{+,+0} \leftrightarrow -{\cal M}_{+,0-}, \quad
   {\cal M}_{-,-0} \leftrightarrow -{\cal M}_{-,0+}.
\end{eqnarray}
The minus sign is due to the fact that $\epsilon^{(1,2)}_0$ 
are obtained from boosting $\hat{z}=(0,0,0,1)$ in the direction of the respective motion of the $Z$.
Therefore under ${\cal R}^\xi$ and P, $\hat{z}\to  - \hat{z}$ and
 $\epsilon_0^{(1)} = \gamma(\beta, 0, 0, 1) \rightarrow  - \epsilon_0^{(2)}$. Moreover the
spin-projection of the parent particle remains unchanged under P.
 One important implication of Eq.~(\ref{rxi}) is there are only two 
independent helicity amplitudes for
any spin-1 particle decaying into two $Z$ bosons. On the other hand, the observation 
that a vector boson is odd under charge conjugation (C) implies all the P-odd amplitudes
should be CP-conserving and real, whereas the P-even amplitudes are CP-violating and purely
imaginary. Therefore we can parametrize the four non-vanishing amplitudes as follows:
\bea {\cal M}_{+,+0} &=&  A + i\, B =  C e^{ i \delta} = -  {\cal M}_{-,0+}, \nonumber\\
     {\cal M}_{+,0-} &=&  A - i\, B  =  C e^{- i  \delta} = - {\cal M}_{-,-0} .
\label{eq:f4f5}
\eea
The parameter $C$ is an overall normalization and will drop out in the 
normalized differential cross section. The phase
$\delta = \tan^{-1}(B/A)$ is 0 for $B=0$ and $\pi/2$ for $A=0$.

To see how 
$\delta$ enters into the angular distributions when $Z_1Z_2$ further decay, 
recall that 
$Z_i$ produces an angular dependence $\exp(i m_i \phi_i)$, where $m_i=\pm,0$ is the 
spin-projection and $\phi_i$ is the 
azimuthal angle in the rest frame of the $Z_i$.
Obviously only the relative angle $\phi$ is physical and we can 
set $\phi_2=0$ and $\phi = \phi_1$. Then  $\delta$ only enters as a phase shift in $\phi \to \phi +2\delta$. 
For example, focusing only on the $\phi$ dependence, 
\begin{equation}
\label{interfere}
 \left|a_1 {\cal M}_{+,+0} e^{i\phi} + a_2 {\cal M}_{+,0-} \right|^2 \sim
                 |a_1 e^{i(\phi+2\delta)} + a_2 |^2,
\end{equation}
and similarly for ${\cal M}_{-,\lambda_1\lambda_2}$. This argument also makes it clear that the 
angular distribution has the 
form
\begin{equation}
\frac{dN}{Nd\phi} \sim c_1 + c_2 \cos (\phi+2\delta).
\end{equation}  
It is worth noting that $\delta$ is the only place where the kinematics of the system matters;
all coefficients
in the differential rate must then be determined by the symmetry. Furthermore, a measurement
on $\delta$ could determine the relative strength 
between the CP-conserving and CP-violating
amplitudes.

\begin{figure}
\includegraphics[scale=0.8,angle=0]{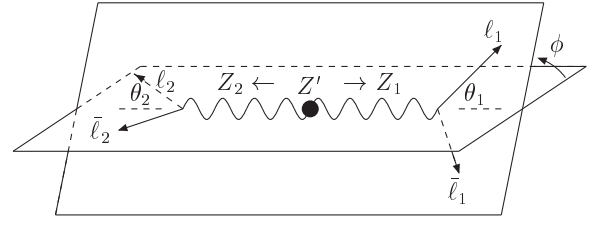}  
\caption{\label{fig1}{\em Two decay planes of $Z_1\to \ell_1\bar\ell_1$  and
$Z_2\to \ell_2\bar\ell_2$ define the azimuthal angle $\phi\in[0,2\pi]$ 
which rotates $\ell_2$ to $\ell_1$ in the transverse view.
The polar angles $\theta_1$ and $\theta_2$ shown are 
defined
in the rest frame of $Z_1$ and $Z_2$, respectively. 
}}
\end{figure}


\noindent {\em Angular Distributions --}
Now we turn to the specific interactions between a $Z'$ and two $Z$ bosons.
 The effective Lagrangian, when all particles are on-shell, includes only two operators at dim-4: 
\begin{equation}
{O}_{CPV} = f_4 Z'_{\mu} (\partial_\nu Z^\mu) Z^\nu,  
{O}_A = f_5 \epsilon^{\mu \nu\rho \sigma} 
Z'_{\mu} Z_\nu (\partial_\rho Z_\sigma) \ .
\end{equation}
In momentum space the form factor for
$Z'(q_1+q_2,\mu) \to Z(q_1,\alpha) Z(q_2,\beta)$ can be written as
\be
\label{Eq:vertex}
\Gamma_{Z'\rightarrow Z_1 Z_2}^{\mu\alpha\beta}
=    i    f_4 (q_2^\alpha g^{\mu\beta} + q_1^\beta g^{\mu\alpha})
       + i f_5  \epsilon^{\mu\alpha\beta\rho}(q_1-q_2)_{\rho}.
\ee 
The non-vanishing helicity  amplitudes are
\bea {\cal M}_{+,+0} &=& - {\cal M}_{-,0+} = R(-f_5\beta  + i f_4) \ ,  \nonumber\\
     {\cal M}_{+,0-} &=& -{\cal M}_{-,-0}= R(-f_5\beta - i f_4) \ ,
\label{eq:f4f6}
\eea
where  $\beta^2=1-4m_Z^2/m_{Z'}^2$ and $R={\beta m_{Z'}^2\over 2 m_Z}$. 
In this case, the phase $\delta = \tan^{-1} ( -f_4/ f_5 \beta )$.

%
%

The helicity state of the $Z$ boson manifests itself in the angular distribution of the
subsequent decay products. Particularly in the purely leptonic decays
$Z_1Z_2 
\rightarrow (\ell \bar\ell)_1 +(\ell\bar\ell)_2$
all kinematics can be measured precisely. The polar angle $\theta_i$ is measured from the direction of motion of
$Z_i$ in the rest frame of $Z'$ to the negatively charged lepton $\ell_i$ in the $Z_i$ rest frame, whereas
the azimuthal angle $\phi$ is measured from $\ell_2$ to $\ell_1$, as illustrated in Fig.~1.
%
In the end the differential cross section can be obtained from the expression,
\bea
\sum_{\kappa,h_1,h_2}
 \left| \sum_{\lambda_1, \lambda_2} 
\mathcal{M}_{\kappa, \lambda_1 \lambda_2}\,
g_{h_1}  f^{h_1}_{\lambda_1}(\theta_1, \phi)\,
g_{h_2} f^{h_2}_{\lambda_2}(\theta_2, 0)\right|^2
\label{eq:distribution}
\eea
where $h$ stands for the chirality ($+,-$ for $R,L$) 
of the lepton with the coupling $g_h$ to the $Z$.  
The spin-1 rotation matrix elements $f_m^h$ are
{
\bea 
\label{eq:zamp}
2 f_{m}^h(\bar{\theta}, \bar{\phi}) 
&=& (1 + m h \cos \bar{\theta} ) {e^{ i m \bar{\phi}}} 
 ,  \nonumber \\
\sqrt{2} f_0^h (\bar{\theta}, \bar{\phi})&=& {h} \sin \bar{\theta}. \eea 
}
Note that $m$ in $f^h_m$ refers to the spin-projection $m=\pm$ and 
$\{\bar{\theta},\bar{\phi}\}$ are defined with respect to the $+z$ axis in
the canonical way. Since $Z_2$ moves in the
$-z$ axis in the rest frame of the $Z'$, and $\theta_2$ is defined from $-\hat{z}$ to $\ell_2$,  
$f_{m_2}^{h_2}$ in Eq.~(\ref{eq:distribution}) should really be 
$f_{-\lambda_2}^{h_2}(\pi-\theta_2,0)$, which equals $f_{\lambda_2}^{h_2}(\theta_2,0)$ 
by Eq.~(\ref{eq:zamp}).
In the end the normalized angular distribution  is
{
\bea
&& {8\pi dN\over N d\cos\theta_1 d\cos\theta_2 d\phi}
={9\over8}\left[
1-\cos^2\theta_1\cos^2{\theta_2} \right.
\nonumber\\
&& \qquad 
-\cos\theta_1\cos{\theta_2}\sin{\theta_2}\sin\theta_1\cos(\phi+2\delta)
 \nonumber\\
&&  \qquad \left.  +\hbox{${ (g_L^2-g_R^2)^2\over(g_L^2+g_R^2)^2 }$}
\sin\theta_1\sin{\theta_2}\cos(\phi+2\delta)\right]  \ . 
\eea
}
We see explicitly that all coefficients in the distribution are independent of kinematic variables 
such as $m_{Z'}$ and $m_Z$; the only dependence on kinematics comes in through the phase shift $\delta$.
As we have seen in the previous section, this is a consequence of the Landau-Yang theorem as the
symmetry of the system completely fixes the numerical coefficients. 
Clearly, the $\phi$ dependence reveals the relative magnitude of the two operators --
a purely anomaly vertex ${O}_A$ 
gives $\delta = 0$ while a purely CP-violating vertex ${O}_{CPV}$ 
has $\delta=\pi /2$. It may seem surprising that we can observe a CP-violating operator without
the presence of CP-conserving interactions, since usually one needs interference
effect to observe CP violation. In our case, it is worth noting that the phase shift in the azimuthal
angle $\phi$ is in fact an interference effect between two different helicity amplitudes, as can be 
seen clearly from Eq.~(\ref{interfere}).

If we integrate over both polar angles $\theta_1$ and $\theta_2$, the $\phi$
dependence is highly suppressed by the approximate relation
$g_L\approx -g_R$ for leptonic decays. The suppression can be understood from
a partial $\hat
C$ symmetry as discussed Ref.~\cite{Chang:1994cs}.
However, if we only integrate polar angles $\cos\theta_1\cos\theta_2>0$
or $<0$, then
\begin{equation} 
{2\pi dN _{\pm}  \over N  d\phi}
={1\over 2}\left[1  +  \left(
 {9\pi^2\over 128}{ (g_L^2-g_R^2)^2\over(g_L^2+g_R^2)^2 }\mp  {1\over 8}\right)
      \cos(\phi+2\delta)\right] 
\end{equation}
where  $N_\pm$ stands for $N(\cos\theta_1\cos\theta_2  {\ }^>_< \ 0)$.
Since $g_L^2 \approx g_R^2$, we will ignore  the $(g_L^2-g_R^2)^2$
term from now on.

We briefly remark that if one or both of the $Z$'s decay hadronically, a larger
sample of events could be collected. In this case a complication arises because
previously we { relied }
on the negatively charged lepton to define the polar and azimuthal
angles. However we could still use the {\em forward jet}, defined as the jet with
polar angle $\theta_i \in[0,{\pi\over2}]$, in place of the negatively charged lepton
and take into account the fact that different kinematic configurations will be in 
experimentally indistinguishable region. In the end we find there is still residual
dependence on the azimuthal angle $\phi+2\delta$. Nonetheless, we anticipate that QCD
background will completely overwhelm the signal in these circumstances 
at the { CERN} LHC. However,
at the linear collider it could be useful to apply the strategy to the hadronic decay
of the $Z$.

\noindent {\em Measurements at the LHC --}
Here we estimate the production rate of the two $Z$'s from the $Z'$ decay that is needed to 
distinguish ${O}_A$ and ${O}_{CPV}$, taking into account background at the LHC. 
Because of the QCD background, we focus on the leptonic decays
$Z' \rightarrow ZZ \rightarrow 4 \ell$. The main background is the SM $ZZ$ production 
$q\bar{q}\to ZZ$ through a $t$-channel diagram which has a cross section of about 15 pb. 
Assuming a small decay width for the $Z'$, 
at the LHC the width measurement will be dominated by the detector energy resolution. 
Based on simulations using 
{ {\small MADEVENT} }\cite{Alwall:2007st}, followed by showering 
and hadronization in 
{ {\small PYTHIA} } \cite{Sjostrand:2006za} 
and detector simulation in PGS4 \cite{PGS},
we find that the $Z'$ width from smearing is about 12 GeV for a 240 GeV mass. 
Therefore we can put an invariant mass cut on the $ZZ$ final state, $234$ GeV $< m_{ZZ} <246$ GeV, to
reduce the SM $ZZ$ background to 79 fb.

If we see a resonance in $ZZ\to 4\ell$ above the SM background at the LHC, the first question is to
determine the spin and CP property of the new particle. Such a question has been studied 
previously in the context of Higgs discovery. We expect that
the spin of the new particle can be determined unambiguously by various 
proposals \cite{Barger:1993wt}. Previous studies, however, assumed a definite 
CP property of the new particle whereas presently we are mainly interested in the CP-hybrid case:
the presence of both ${O}_A$ and ${O}_{CPV}$. For a $5\sigma$ discovery of $Z'$ through
its decay to $ZZ\to 4\ell$, 
we need the 
ratio of the signal $S$ to the statistical error in the background $\sqrt{B}$ to be 5. 
For 100 fb$^{-1}$ luminosity at the LHC, 
we find the required production for $ZZ$ is 67 fb for a 240 GeV $Z'$. 



The next step after the discovery is measurement of the azimuthal angular distribution and the phase
shift $\delta$. If we include the SM background and assume it has a flat distribution , the expected distribution becomes 
\bea n_\pm(\phi) \equiv {dN_\pm
\over d\phi}
={N \over 4 \pi}\left[
1 \mp \frac18 \frac{S}{S+B} \cos(\phi+2\delta)  \right]  . 
 \eea
A Bayesian method can be used o fit the phase $\delta$ with its
central value and statistical error. At this stage we will be content
with a simple event counting to estimate the required production rate
of $Z'$ in order to distinguish ${O}_A$ from the operator ${O}_{CPV}$.
We define an ``up-down asymmetry" { $\mathcal{A}_{ud}$ } in the
absence of background as {
\begin{equation}
 {\left( \int_{-\pi/2}^{\pi/2} - \int_{\pi/2}^{3 \pi/2} \right) \frac{ n_{+}(\phi) - n_{-}(\phi)}{N} d \phi } 
= - {\cos ( 2 \delta ) \over 4 \pi}.
\end{equation}
}
For operator ${O}_A$ only ($\delta = 0$), {
$\mathcal{A}_{ud} = -{1\over  4\pi}$, 
whereas $\mathcal{A}_{ud} = {1\over4\pi}$}
for ${O}_{CPV}$ only ($\delta = \pi /2$). 
If we would like to discriminate ${O}_A$ from ${O}_{CPV}$ at 99.7\% 
confidence level (3$\sigma$), 
we would require the difference in the number of the ``up-down'' asymmetrical events 
$S_A = {\cal A}_{ud}\times S$
over the statistical errors of
 the total number of events $\sqrt{S + B}$ to be 3:
\begin{equation}
\frac{\left| S_A(\delta=0)-S_A(\delta=\pi/2)\right|}{\sqrt{S+B}} = {S \over 2 \pi \sqrt{S+B}} 
              = 3 \ .
\end{equation}
Then the required production rate of the $Z$ boson from $Z'$ decay is 0.9 pb for a 240 GeV $Z'$.

Such a production rate could in fact be easily fulfilled in models where the branching 
ratio of $Z' \rightarrow ZZ$ is not very small. For instance, in the littlest Higgs model with 
anomalous T parity, the lightest T-odd particle $B_H$ (the $Z'$) will only decay into
$W^+W^-$ and $ZZ$ through the 
WZW term, which gives rise to a non-zero ${O}_A$, and
the branching ratio of $Z' \rightarrow ZZ$ turns out to be roughly 
1/3 \cite{Barger:2007df}. 
Since the T-odd particle is always pair-produced and eventually 
decays into  $B_H$, the required $B_H$ production rate, which is also
the total cross section for T-odd particles, will be 1.3 pb. 
If we choose a typical parameter $f = 1.5$ TeV, $m_{B_H} = 240$ GeV, 
we find that even the T-quark production channel alone with a T-quark mass 
750 GeV will give us the necessary production rate for discrimination \cite{Carena:2006jx}.

\noindent {\em Conclusion --}
{ 
In this work we study decays of a new $Z'$ gauge boson into two SM $Z$
bosons by extending the Landau-Yang theorem to a parent particle
decaying into two $Z$ bosons. The original theorem forbids decays of a
spin-1 particle into two photons based on simple symmetry arguments.}
{ We show the generalized Landau-Yang theorem  makes} strong
predictions on the polarization states of the two $Z$'s and the
differential distribution of its subsequent decays. In particular,
the normalized differential cross section depends on kinematics only
through a phase shift in the azimuthal angle between the two decay
planes of the $Z$. For a $Z'$ boson, the two couplings are
anomaly-induced and CP-violating, whose effects could be disentangled
by measuring the phase shift at the { CERN} LHC.
The method could potentially be applied to the case of a 
scalar decaying into $ZZ$ final state
which does not have a definite CP property.


\noindent {\em Acknowledgements --}
This work was supported in part by the U.S. Department of Energy under
DE-FG02-84ER40173 (W.-Y.~K.) and 
DE-AC02-06CH11357 (I.~L. and J.~S.). J.~S. was also supported
by the University of Chicago under section H.28 of its contract 
W-31-109-ENG-38 to manage Argonne National Laboratory.
W.-Y.~K. also thanks NCTS, Taiwan for hospitality.

\end{document}